\documentclass[aps,prl,showpacs,showkeys,floatfix,nofootinbib,twocolumn, notitlepage,superscriptaddress]{revtex4-1}  
\usepackage{graphicx}  
\usepackage{subcaption}
\usepackage{dcolumn}   
\usepackage{bm}        
\usepackage{braket}
\usepackage{exscale}                  
\usepackage[intlimits]{amsmath}       
\usepackage{amsfonts}
\usepackage{amssymb,amscd}
\usepackage{color}
\usepackage{hyperref}
\usepackage{slashed}
\usepackage[normalem]{ulem}
\usepackage{graphicx}
\usepackage{multirow}
\usepackage{float}
\usepackage{comment}

\newcommand{\xs}{{\bf x}}

\newcommand{\ps}{{\bf p}}
\newcommand{\qs}{{\bf q}}

\begin{document}

\title{Heavy Quark Momentum Broadening in a Non-Abelian Plasma away from Thermal Equilibrium} 

\author{Harshit Pandey}
\email{harshitp@imsc.res.in} 
\affiliation{The Institute of Mathematical Sciences, a CI of Homi Bhabha National Institute, Chennai 600113, India}
\author{S\"{o}ren Schlichting}
\affiliation{Fakult\"{a}t f\"{u}r Physik, Universit\"{a}t Bielefeld, Bielefeld D-33615, Germany}
\author{Sayantan Sharma}
\affiliation{The Institute of Mathematical Sciences, a CI of Homi Bhabha National Institute, Chennai 600113, India}

\begin{abstract}  
We perform classical-statistical real-time lattice simulations to compute real-time spectral 
functions and momentum broadening of quarks in the presence of strongly populated non-Abelian 
gauge fields. Based on a novel methodology to extract the momentum broadening for relativistic 
quarks, we find that the momentum distribution of quarks exhibit interesting non-perturbative 
features as a function of time due to correlated momentum kicks it receives from the medium, 
eventually going over to a diffusive regime. We extract the momentum diffusion coefficient 
for a mass range describing charm and bottom quarks and find sizeable discrepancies from 
the heavy quark limit.
\end{abstract}

\maketitle

\textit{Introduction}
Heavy quarks are excellent probes of the Quark Gluon Plasma (QGP) formed in collisions (HIC)
of heavy nuclei, performed in large-scale collider experiments at RHIC and LHC~\cite{Rapp:2018qla}. 
Heavy charm (c) and  bottom (b) quarks are particularly interesting since these are almost 
exclusively formed in the very early stages of ultra-relativistic HICs through perturbative 
hard scattering process of the colliding nuclei.  Subsequently, these heavy quarks interact
with the QGP formed in HICs, thus charting out its entire time history until hadronization. 
Historically, the in-medium dynamics of heavy quarks has primarily been modeled as a Brownian 
motion in a medium consisting of colored degrees of freedom through the Langevin 
equations~\cite{Svetitsky:1987gq,Moore:2004tg}. More recently,  the effect of 
color interactions~\cite{Liu:2019lac} have been implemented including color memory 
effects~\cite{Ruggieri:2022kxv,Ruggieri:2019zos} in the Langevin dynamics. Despite 
the existence of successful phenomenological models, major theoretical challenges 
in the description of heavy quark dynamics still remain to be addressed. Firstly the 
QCD medium traversed by the heavy quark both in the pre- and near-equilibrium stages 
interacts with it non-perturbatively~\cite{vanHees:2004gq,vanHees:2007me,Akamatsu:2008ge,Akamatsu:2012vt}, 
and gluon (color) exchange processes cannot be described semi-classically except in the 
limit of infinite number of colors~\cite{Blaizot:2017ypk}. Because of this the dynamics of 
heavy quarks in a colored medium can be described using Langevin equation only under specific 
circumstances~\cite{Blaizot:2017ypk}.
Secondly, in view of the experimental data on the elliptic flow $v_2$ for charm quarks~\cite{STAR:2006btx, 
PHENIX:2006iih,ALICE:2017pbx,STAR:2017kkh} -- the non-relativistic approximation often applied for the 
heavy quarks may not be well justified for charm quarks. Since the experimentally observed $v_2$ of 
charm quarks is consistent with that of light quarks, this may hint towards the need for a full 
relativistic treatment of charm quark dynamics beyond the existing studies at or close to the 
infinite-mass limit~\cite{Zhang:2005ni}.

To understand the interactions of the heavy quarks with the QCD medium, a quantity of interest 
is the heavy quark diffusion coefficient $\kappa$, which quantifies the auto-correlation function 
of the noise term that model the random momentum kicks that it receives in the medium during its 
evolution~\cite{Braaten:1991jj, Braaten:1991we}. In a plasma in local thermal equilibrium, the 
fluctuation-dissipation theorem relates $\kappa$ to the drag coefficient which quantifies 
the rate at which kinetic equilibration occurs. Since its reconstruction at the level of spectral 
function is difficult~\cite{Petreczky:2005nh}, at leading order in $1/m$, where the mass $m$
of the heavy quarks which is assumed to be much larger than all other scales, the coefficient 
$\kappa$ can be related to the spectral function of the color-electric field correlator~
\cite{Caron-Huot:2009ncn,Casalderrey-Solana:2006fio}. In this approximation, $\kappa$ has 
been calculated within perturbative QCD, showing a poor convergence~\cite{Laine:2009dd} emphasizing 
the need for its non-perturbative calculation. Such a non-perturbative calculation in the continuum 
limit using lattice techniques have been performed both in the quenched 
case~\cite{Laine:2009dd,Meyer:2010tt,Francis:2011gc,Banerjee:2011ra,Ding:2018uhl, Francis:2015daa, Altenkort:2020fgs,Brambilla:2020siz} 
as well as in QCD medium with dynamical fermions~\cite{Altenkort:2023oms}. These results on $\kappa$ are 
significantly lower than the perturbative estimates~\cite{Caron-Huot:2007rwy,Caron-Huot:2008dyw} 
moving towards the upper edge of the theoretical systematic uncertainty band of weak-coupling calculations
at temperatures of $\sim 1$ GeV~\cite{Brambilla:2020siz}. 
Very recently $1/m^2$ corrections to $\kappa$ have been calculated, which receive different 
contributions, significant among which is the color magnetic part of the Lorentz 
force~\cite{Bouttefeux:2020ycy}. The $\kappa$ that arises out of the color magnetic 
correlator has non-trivial renormalization effects which are being investigated only in recent years~\cite{Banerjee:2022uge, Banerjee:2022gen, Altenkort:2023eav,Brambilla:2022xbd,Brambilla:2023vwm}. 
The magnetic part of $\kappa$ is comparable to its electric counterpart~\cite{Brambilla:2022xbd}, which 
implies a $20~(34)\%$ correction to $\kappa$ for bottom (charm) quarks respectively using lattice results 
of their mean-squared velocities~\cite{Petreczky:2008px}.

Recent phenomenological studies also demonstrate that a significant amount of $v_2$ for heavy quarks 
already starts to build up during the pre-equilibrium stages of HIC~\cite{Sun:2019fud,Song:2019cqz}, 
hence a precise determination of $\kappa$ in this regime is important. At early times this medium has 
non-perturbatively high occupation numbers of low-momentum gluons, hence the precise determination 
of $\kappa$ is challenging. The $\kappa$ has been estimated within transport 
models~\cite{Mrowczynski:2017kso,Cao:2018ews,Rapp:2018qla,Du:2023izb} as well as in classical-statistical 
lattice gauge theory simulations in the infinite quark mass 
limit~\cite{Boguslavski:2020tqz,Avramescu:2023qvv} showing
a sizeable difference from perturbative kinetic theory estimates~\cite{Boguslavski:2023fdm}. 
It will be thus desirable to develop a  more fundamental theoretical formalism to measure 
$\kappa$ without resorting to an expansion in powers of $1/m$. 
We thus, for the first time report the value of the heavy quark diffusion coefficient in a 
non-equilibrium plasma, treating them as relativistic particles. We choose a specific 
non-equilibrium condition of a highly occupied non-Abelian plasma in a self-similar scaling 
regime at its classical fixed point and perform a time evolution of quarks inside such a medium. 
We monitor the momentum broadening as a function of time and measure its $\kappa$ from the width 
of the distribution. Our formalism is very general which enables us to measure $\kappa$ for a wide 
range of bare quark masses.

\textit{Details of the classical-statistical lattice simulations}
We follow previous works~\cite{Kurkela:2012hp, Berges:2013fga, Berges:2013eia, Schlichting:2022fjc} 
and first generate gauge field configurations using a classical-statistical evolution of the color fields 
using the classical Yang-Mills equations of motion defined using the (Minkowski) Wilson gauge action on a lattice of spatial volume $N^3$ and spacing $a$.
Color electric fields $\mathbf{E}^a_{i}$ and gauge links $U_{i}$ are time-evolved up to a reference 
time of $Qt=500, 1000, 1500$ using a leap-frog integrator, where $Q$ is the characteristic hard 
scale in the problem. By these times the gauge field evolution reaches a self-similar regime where the 
momentum distribution function $f_g$ attains a fixed point scaling with time, of the form 
$g^2 f_g(|\mathbf{p}|,t)=(Qt)^{-\frac{4}{7}}~f_s\left[(Qt)^{-\frac{1}{7}}\frac{|\mathbf{p}|}{Q}\right]$ with a scaling 
function $f_s$~\cite{Berges:2013fga,Berges:2013eia}.
We perform our simulations with two different initial conditions for the $SU(2)$ gauge group for performing statistical averages, and if not stated otherwise we consider lattices 
of size $N=256$ with lattice spacing $a_s=0.5/Q$ and a time step $a_t=0.025~Q$ in our work.

Over the course of the self-similar evolution, the non-equilibrium plasma establishes a separation of 
scales~\cite{Berges:2017igc,Berges:2023sbs}, which is analogous to the separation of scales in a high-
temperature equilibrium plasma. The hardest scale is set by the highest momenta 
$\Lambda(t) \sim Q (Qt)^{1/7}$, the electric screening (Debye) scale is $m_D(t) \sim Q (Qt)^{-1/7}$ 
and the magnetic screening scale is simply $\sqrt{\sigma(t)} \sim Q (Qt)^{-3/10}$.  While at 
initial times all these scales are $\sim Q$, they start to separate dynamically 
culminating in the scaling regime where $\sqrt{\sigma(t)} < m_D(t) \ll \Lambda(t)$, 
such that at  $Qt=1500~(1750)$, where we typically start (end) our measurements, the 
scales are $\Lambda~= 2.1~Q~(2.14~Q),~m_D~= 0.21~Q,~(0.206~Q)$ and $\sqrt{\sigma}~= 0.03~Q,~(0.029~Q)$ 
respectively~\cite{Boguslavski:2018beu}.

We then study the dynamics of relativistic quarks in this scaling regime, as described in 
detail in Appendix A. Starting from a reference time $Qt$, we initialize the 
fermionic wavefunction as $\phi^{u/v}_{\lambda,s}(t,\xs)=u/v_{\lambda,s}({\bf P}) e^{\pm i{\bf P}.{\bf x}}$.
\footnote{We globally employ temporal axial ($A^{0}=0$) gauge, and fix the residual gauge freedom by 
further implementing the Coulomb gauge condition ($\partial_{i}A^{i}=0$) at the time of initialization 
and measurement of gauge dependent observables~\cite{Berges:2013fga,Boguslavski:2021kdd}.}, representing 
a free relativistic quark/anti-quark with a fixed momentum $\mathbf{P}$, spin index $s=1,2$ and color 
index $\lambda=1,\cdots,N_c$ at the initial time $Qt$. Subsequently, the quark wave functions 
$\phi^{u,v}_{\lambda,s}(t,\xs)$ are then evolved quantum-mechanically as a function of time, $t'$, 
in the background of classical color gauge fields using the $\mathcal{O}(a^2)$-improved Wilson-Dirac 
Hamiltonian on the lattice,
\begin{equation}
i\partial_{t'} \phi^{u,v}_{\lambda,s}(t',\xs)=\gamma_0\left[-i\gamma_i D_i+m\right] 
\phi^{u,v}_{\lambda,s}(t',\xs)~.
\end{equation}
where $D_i (i=1,2,3)$ represents the spatial part of the Wilson-Dirac operator in presence of the 
gauge fields.
By performing a spatial Fourier transform of the time-evolved quark wavefunctions at time 
$t^\prime>t$ denoted as $\Tilde{\phi}^{u,v}_{\lambda,s,\ps}(t^\prime, \ps)$, the quark spectral 
function can be calculated by projection on the free Dirac spinors corresponding to (anti) 
particles $(v_{\lambda,s} (\ps))~u_{\lambda,s} (\ps)$ as~\cite{Boguslavski:2021kdd}
\begin{eqnarray}
    &&\rho(t^\prime, \ps)^{\alpha \beta}= \\
    &&\frac{1}{N^3} \sum_{\lambda,s} \Big[  \Tilde{\phi}^{u,\alpha}_{\lambda,s}(t^\prime, \ps) u^{\dagger,\delta}_{\lambda,s} (\ps) +  
    \Tilde{\phi}^{v,\alpha}_{\lambda,s}(t^\prime, \ps) v^{\dagger,\delta}_{\lambda,s} (-\ps)  \Big] \gamma^{\delta \beta}_0~, \nonumber
\end{eqnarray}
which can then be decomposed into scalar ($\rho_S$), pseudo-scalar ($\rho_P$), vector ($\rho^\mu_V$), 
axial-vector ($\rho^\mu_A$) and tensor ($\rho^{\mu \nu}_T$) components using the Clifford basis.

Similarly, the momentum distribution $dN/d^3{\bf q}$ at time $t^\prime= t+\Delta t$ is calculated by 
projecting the evolved quark wave functions $\Tilde{\phi}^u_{\lambda,s}(t^{\prime},\mathbf{x})$ 
(measured in the Coulomb gauge) onto the non-interacting eigenstates 
$u^\dagger_{\lambda',s'}(\mathbf{q})e^{-i{\bf q}.{\bf x}}$ according to
\begin{eqnarray}
    \frac{dN}{d^3 \mathbf{q}}= \frac{1}{2N_c}\sum_{\lambda,\lambda',s,s'} 
    \vert u^\dagger_{\lambda',s'}(\mathbf{q}) \Tilde{\phi}^u_{\lambda,s}(t^{\prime},\mathbf{q})\vert^2~.
\end{eqnarray}
where the sums over $\lambda,\lambda',s,s'$ correspond to averaging over initial and final spin and color 
states. Since for initial momentum ${\bf P}\equiv(0,0,0)$, the momentum distribution $dN/d^3 {\bf q}$
measures the momentum acquired by the quark through interactions with the medium, the second moment 
of the distribution can then used to calculate the diffusion coefficient $\kappa$ as discussed further 
in the penultimate section (also see Appendix B).

\textit{Quark spectral functions -- Effective masses and lifetimes} 
We first investigate the behavior of the spectral function of a quark inside the medium, 
which contains information about its dynamics in an interacting quantum field theory. We 
will be considering a wide range of bare quark masses $m_{\rm bare}=m$ in units of the initial 
hard scale $Q$ and study the corresponding spectral functions inside the highly-occupied 
non-Abelian plasma described in the previous section. We anticipate that in the non-equilibrium 
Glasma created in the initial stages of high-energy heavy-ion collisions the characteristic 
hard momentum scale $\Lambda \gtrsim 1~\text{GeV}$, such that the values $m/Q=1$-$2$ and 
$m/Q=4$-$8$ represent quarks with masses close to charm and bottom quarks respectively.

\begin{figure}[hb]
    \centering
    \raisebox{-\height}{\includegraphics[width=0.45\textwidth]{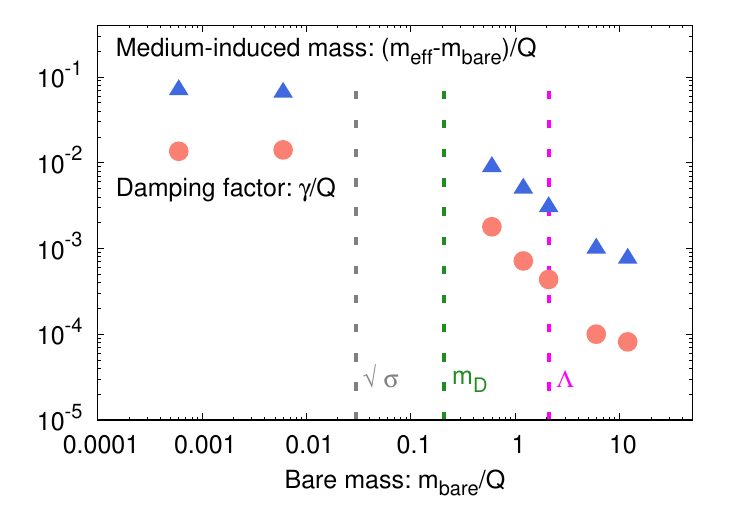}}
    \caption{Dependence of the medium-induced mass $m_{\rm eff}-m_{\rm bare}$ (triangles) 
    and the damping factor $\gamma$ (circles)  on the bare quark mass $m_{\rm bare}$ at 
    $Qt=1500$. Vertical lines indicate magnetic ($\sqrt{\sigma}$) and electric $(m_D)$ screening scales, 
    as well as the hard scale $\Lambda$.}
    \label{fig:MediumInd}
\end{figure}

We recall that in a previous study~\cite{Boguslavski:2021kdd}, it was shown that the behavior of the 
spectral function for massless quarks can be well described by a modification of the hard-thermal loop 
(HTL) result~\cite{Blaizot:2001nr} introducing an effective damping rate $\gamma$, such that for vanishing 
spatial momentum $\vert {\bf p}\vert =0$, the vector component $\rho_V$ of the spectral function 
can be parametrized as~\cite{Boguslavski:2021kdd}, 
$\text{Re }\rho_{V}^{0} (t) = \rm{e}^{-\gamma.t} \cos (m_{\text{eff}}.t)$.
We find that this ansatz also describes the spectral functions of heavy flavor quarks, when the 
medium induced quark mass $m_{\rm eff}$ and damping rate $\gamma$ are considered to depend on the 
bare quark mass $m_{\rm bare}$. Only for intermediate quark masses, 
$(m_{\rm eff}-m_{\rm bare}) \sim m_{\rm bare}$, we find that the spectral function is not well described by this rather 
simple form (see Appendix C). By performing a fit (c.f. Appendix C for details) of the above 
ansatz to the real-time spectral functions in a time interval $Q \Delta t=500 (100)$ for bare 
quark masses $m_{\text{bare}}/Q<0.01 ~(\geq 0.6)$, we obtain the medium induced masses and the 
widths shown in Fig.~\ref{fig:MediumInd}.  The medium 
induced mass $m_{\text{eff}}-m_{\text{bare}}$ is $\sim 0.08~Q$ for a range of bare masses for which $m_{\text{bare}}< (m_{\text{eff}}-m_{\text{bare}})$ beyond which it decreases rapidly falling to less than a 
tenth of that value for $m_{\text{bare}}=2.1~Q \sim \Lambda(t)$. This suggests that for the 
typical mass range near the bottom quarks the medium-induced effects are negligibly tiny 
whereas for the charm quarks, it can still be significant but less than that of the light 
quarks. As a naive comparison, for $2+1$ flavor QCD in thermal equilibrium, the thermal mass for 
light quarks comes out to be $m_{\text{th}}/T=0.725(14)$ at around $T=3T_c\simeq 470$ 
MeV~\cite{Karsch:2009tp}, though the spectral functions in both these regimes 
can be very different and needs a dedicated study~\cite{Beraudo:2010tw,Boguslavski:2021kdd}.

From the plot of the damping factor $\gamma$ as a function of the bare quark mass in 
Fig.~\ref{fig:MediumInd}, we conclude that the width of the effective mass peak
in the spectral function is quite large $\sim 20\%$ for the range of $m_{\text{bare}}$ 
corresponding to the light quarks, falling to less than $\sim 10\%$ for heavier quarks. 
Importantly, if the width is too broad, a heavy quark in the medium will 
correspond to a collective excitation and cannot be simply treated as a weakly interacting 
quasi-particle. Since the typical spectral width of heavy quasi-particles is significantly smaller 
than the lighter counterparts, these can be treated as well-defined quasi-particles and their 
dynamics can be modelled in terms of a Langevin equation~\cite{Svetitsky:1987gq}.  Conversely, 
for light quarks one would require a more involved treatment of the medium-particle interactions, 
as their large width of the spectral function indicates a significant modification due to the medium.  
Interestingly, particles with bare mass of the same order as the charm quark, have intermediate widths 
between the very heavy and light quarks and thus require a careful treatment, beyond the infinite-mass 
approximation.

\begin{figure}[b]
    \centering
    \raisebox{-\height}{\includegraphics[width=0.48\textwidth]{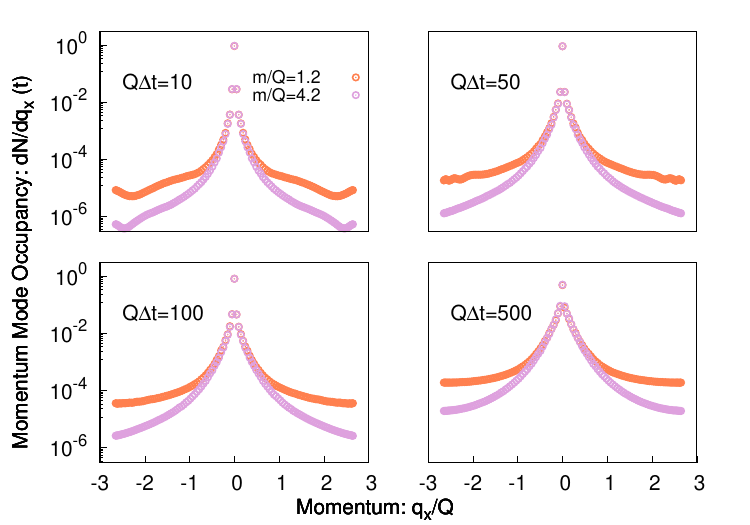}}
    \caption{Evolution of the momentum  distribution $dN/d q_x$  as a function of time $Q \Delta t$ for quark 
    masses $m/Q=1.2$ and $m/Q=4.2$. Starting from vanishing momentum $dN/d q_x=\delta( q_{x})$ at initial time 
    $Qt=1500$, the quarks acquire momentum through interaction with the non-Abelian plasma.}
    \label{fig:mmod1.2v2.4Q}
\end{figure}

\textit{Momentum broadening of heavy quarks} We next study the scenario where a heavy quark with 
some fixed initial momentum propagates through the medium. It is expected that if the quark behaves 
like a Brownian particle and experiences uncorrelated kicks from the gluons, it will diffuse with time 
and its momentum distribution will show a spread about its initial momentum value which can be quantified 
though the heavy quark momentum diffusion coefficient $\kappa$. By analyzing the momentum spectrum 
$dN/d^3{\bf q}$ of a quark evolving in the dense gluonic medium for different values of the bare quark 
mass $m$, we can then investigate to what extent the dynamics of a heavy quark can be 
approximated in terms of the motion of a non-relativistic Brownian particle.

We present a compact summary of our findings in Fig.~\ref{fig:mmod1.2v2.4Q},  where we show the 
momentum distribution along the $q_x$-direction (obtained by integrating over spatial momenta 
$q_y$ and $q_z$) at different times $Q \Delta t$  as a function of the $\mathcal{O}(a^2)$
improved lattice momentum $\hat{q}_x= - 4/3\sin ( 2\pi n_x/N)+1/6 \sin ( 4\pi n_x/N)$ for 
quark masses $m/Q=1.2, 4.2$. Starting with an initial distribution which is a $\delta$-function 
at momentum $\mathbf{q}\equiv (0,0,0)$, we observe that the momentum broadening is clearly dependent 
on the quark mass, and larger for the lighter quark mass i.e $1.2~Q$ compared to $4.2~Q$. Significant 
differences emerge in the tails of the momentum distribution, which corresponds to momentum transfer 
due to hard scattering with the medium constituents, while the broadening of the central peak is 
quite similar for the different masses.

\begin{figure}[ht]
    \centering
    \raisebox{-\height}{\includegraphics[width=0.45\textwidth]{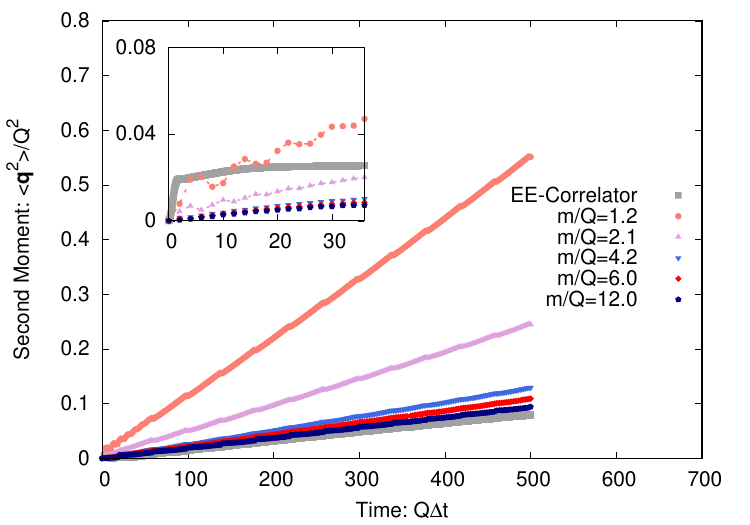}}
    \caption{
    Evolution of the second moment $\langle \mathbf{ q}^2\rangle$ of the momentum distribution 
    $dN/d^3{\bf q}$ as a function of the propagation time $\Delta t$ in the medium. Colored curves 
    indicate the results for different bare quarks masses,  while the gray curve shows the result 
    obtained in the infinite mass limit from the color-electric field two-point correlation function, as 
    in~\cite{Boguslavski:2020tqz}. Differences between the relativistic quantum mechanical treatment and 
    the effective description at early times, are highlighted in the inset.}
    \label{fig:SecondMomPlot}
\end{figure}

We further quantify the momentum broadening through the second moment of the momentum distribution, 
which measures the effective width of the distribution and hence 
the variance of the momentum along $q_x$: Our results presented in Fig.~\ref{fig:SecondMomPlot} 
show distinct features as a function of time; first a transient oscillatory behavior due to the 
initial drift in the correlated gluonic medium and then a subsequent monotonic increase with time. 
In the same Fig. \ref{fig:SecondMomPlot}, the inset shows a comparative study of the momentum 
broadening at very early times, for a range of quark masses and also in the static i.e. 
$m\gg \Lambda(t)$ limit, shown as a gray line. 
The $\langle \mathbf{q}^2 \rangle$ calculated in the static limit as obtained from the color-electric 
field two-point correlator shows a very rapid early-time growth $\propto (\Delta t)^2$, before 
settling to an approximately linear rise $\propto \Delta t$ at later times. By following the 
arguments of~\cite{Boguslavski:2023fdm}, the initial quadratic rise occurs due to the fact 
that the high energy ($\sim \Lambda(t)$) classical gauge field modes act coherently for very 
initial times $\Delta t \lesssim \Lambda(t)^{-1}$. Even though a classical particle would feel 
the this coherent force, the quarks being inherently quantum mechanical excitations, do not 
feature this coherent early-time behavior and instead only exhibit the approximately linear 
rise, at later times where they can effectively be treated as classical colored particles.
By comparing the late time momentum broadening $\langle \mathbf{q}^2 \rangle$ for different 
masses $m$ in Fig.~\ref{fig:SecondMomPlot}, we also observe that for $m/Q\geq 6$, the results 
converge towards the infinite mass limit given by the color electric field correlator. However, 
in the mass range of charm and bottom quarks there are still significant differences which 
highlights the importance of inclusion of relativistic effects, especially for the lighter 
charm quarks.

\begin{figure}[ht]
    \centering
    \raisebox{-\height}{\includegraphics[width=0.48\textwidth]{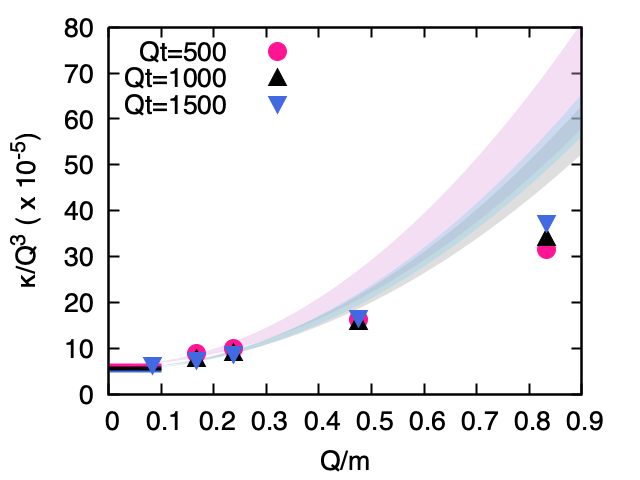}}
    \caption{The momentum diffusion coefficient $\kappa$ as a function of inverse of bare quark mass, 
    at different times $Qt=500,1000,1500$. Solid lines at the bottom left corner correspond to the 
    values obtained in the infinite-mass limit using the color-electric correlator. Shaded bands 
    represent $1/m^2$ correction to the infinite-mass limit. }
    \label{fig:MassVsKappa}
\end{figure}

Since at $Qt=1500$ the evolution of the non-equilibrium plasma is slow compared to the time 
scale of the measurement, a fit to the momentum broadening data at late times $Q\Delta t> 250$ 
for quark masses $m/Q\geq 1.2$ shows a linear growth as a function of time. Such a linear growth 
then signifies the onset of a diffusive regime, and we can therefore extract the momentum diffusion 
coefficient $\kappa$ from the slope of the linear fit to $\langle \mathbf{q}^2 \rangle$ at late times. 
For $m/Q=1.2~(2.1)$, we extract a $\kappa=37.13(5)\times 10^{-5}Q^3~(16.54(4) 
\times 10^{-5}Q^3)$ which shows a significant departure when compared to the earlier estimate 
of $\kappa=5.9\times 10^{-5}Q^3$ obtained in the infinite mass limit~\cite{Boguslavski:2020tqz}. 
On the other hand $\kappa=8.64(2)\times 10^{-5}Q^3~(7.24(1) \times 10^{-5}Q^3)$ for the 
bare quark mass $m/Q=4.2~(6.0)$ similar to the bottom quark is much closer to the static estimate.
The results for the variation of the momentum diffusion coefficient as a function 
of the inverse of bare quark mass in units of $Q$, at different times of initiation 
of quark motion $Qt=500$-$1500$, are summarized in Fig.~\ref{fig:MassVsKappa}. 
The lines in the bottom left corner of the plot are the values of $\kappa$ obtained in 
the infinitely massive limit and agree well with the calculation in~\cite{Boguslavski:2020tqz}. 
To quantify whether this deviation in the range of masses between charm and bottom quarks can 
be understood in terms of $1/m^2$ correction to the static limit estimate, $\kappa(\infty)$, 
we perform a fit of the values of $\kappa(m)/Q^3=\kappa(\infty)/Q^3+\mathcal{O}(1/m^2)$ for 
$m/Q>4.2$ where this ansatz is appropriate, and then extrapolate to lower values of $m$. If 
charm, bottom quarks also follow this trend then the $\kappa(m \leq 4.2)$ will lie on the 
extrapolated graphs, which is not observed in our data, re-emphasizing the need for a relativistic 
treatment of heavy-flavor dynamics.

\textit{Summary \& outlook}
Based on microscopic ab-initio lattice simulations we show for the first time 
the detailed dynamics of a relativistic quark in a highly occupied
non-Abelian plasma. We propose a new method to measure the momentum distribution of 
a quark traversing in the plasma, and based on this method report a significant excess of the momentum 
diffusion of quarks in the mass range from charm to bottom, than previously calculated 
in the static limit, thus highlighting the need for treating their dynamics through a relativistic 
Dirac equation. Beyond the corrections to the infinite mass limit, we also observe significant correlations 
through the interaction of quarks with the medium, which manifest themselves in oscillations of the 
$\langle{\bf q}^2\rangle(\Delta t)$, indicating that a Markovian description is only applicable on 
relatively large time scales. 

Since a highly occupied gluonic plasma considered in our analysis is believed to be produced in the initial 
stages of heavy-ion collisions, our study re-emphasizes the fact that a significant momentum broadening 
and hence a build-up of elliptic flow of charm and bottom quarks may already occur in the initial stages 
of heavy-ion collisions~\cite{Sun:2019fud,Song:2019cqz,Avramescu:2023qvv,Boguslavski:2023fdm}. However, 
to unambiguously confirm this, our calculation needs to be extended to an expanding space-time geometry 
to properly capture the dynamics of the correlated gluon fields in the Glasma. Since our formalism for
investigating the dynamics of quark probes is quite general, one can also study the dynamics of 
light-quark jets through a highly-occupied non-Abelian plasma, which we would like to address 
in a future study.

\textit{Acknowledgements.}
This work is supported in part by the Deutsche Forschungsgemeinschaft (DFG, German Research Foundation) 
through the CRC-TR 211 ’Strong-interaction matter under extreme conditions’– project number 
315477589 – TRR 211. Sa.S. gratefully acknowledges partial support from the Department of 
Science and Technology, Govt. of India through a Ramanujan fellowship when a major portion 
of this work was performed and from the Institute of Mathematical Sciences. The authors acknowledge computing time provided by the Paderborn Center for Parallel Computing (PC2).

\subsection{Appendix A: Details of the momentum mode occupancy distribution} 
\label{app:A}
When investigating the dynamics of heavy and light flavor quarks, we start with a single quark in a fixed momentum mode labelled by momentum 
$\mathbf{P}$, spin polarization $s$ and color $\lambda$ and let it evolve in the background of gauge fields 
in the self-similar regime up to some time $t^{\prime}$. By following the methodology of \cite{Mace:2016shq,Mace:2016svc,Mace:2019cqo} the associated fermionic field operator $\Psi(t',\xs)$ can be 
defined in terms of the time-independent operators $b^{\dagger}_{\lambda,s}(t=0,\ps)$ and 
$d^{\dagger}_{\lambda,s}(t=0,\ps)$ and time dependent wavefunctions 
$\phi_{\lambda,s}^{u,v}$ as,
\begin{eqnarray}
        \Psi(t^{\prime},\xs)= \frac{1}{\sqrt{N^{3}}} \sum_{\lambda,s, \ps} \Big[  \phi^u_{\lambda,s}(t^{\prime},\xs) b_{\lambda,s}(t=0,\ps) + \nonumber \\ \phi^v_{\lambda,s}(t^{\prime},\xs) d^{\dagger}_{\lambda,s}(t=0,\ps)  \Big]~,
\end{eqnarray}
where for our initial conditions the creation/annihilation operators satisfy 
\begin{eqnarray}
    \langle b_{\lambda,m}^\dagger(t=0,\ps) b_{\lambda^\prime,n}(t=0,\ps^\prime) \rangle &=& 
    \delta_{\lambda,\lambda^\prime} \delta_{m,n} \delta_{\ps,\ps^\prime} 
    \delta_{\ps^\prime,\mathbf{P}}  \nonumber \\
    \langle d_{\lambda,m}^\dagger(t=0,\ps) d_{\lambda^\prime,n}(t=0,\ps^\prime) \rangle &=& 0~,
    \label{eqn:initial_cond}
\end{eqnarray}
Now in order to extract the momentum distribution of quarks inside the medium, we make use of the fact that at any time $t$, the fermion field operator $\Psi(t,\xs)$  can be decomposed in terms of time-evolved creation and annihilation operators, labeled by the momentum $\ps$, the color index $\lambda$ and spin index $s$ , as 
\begin{eqnarray}
        \Psi(t,\xs)= \frac{1}{\sqrt{N^{3}}} \sum_{\lambda,s,\ps} \Big(  u_{\lambda,s}(\ps)~ b_{\lambda,s}(t,\ps) \rm{e}^{-i\mathbf{p.x}} + \nonumber \\ v_{\lambda,s}(\ps) d^{\dagger}_{\lambda,s}(t,\ps) \rm{e}^{+i\mathbf{p.x}}  \Big)~.
\end{eqnarray}
By inverting the above expression using the orthonormality of Dirac spinors, we can obtain then the 
time-dependent creation operators $b_{\lambda^\prime,s^\prime}(t,\mathbf{q})$ in terms of Fourier 
transform of the time-evolved fermion fields as,
\begin{eqnarray}
        b_{\lambda^\prime,s^\prime}(t,\mathbf{q}) &=& \sum_{\xs} u^{\dagger}_{\lambda^\prime,s^\prime}(\mathbf{q}) \Psi(t,\xs) ~\rm{e}^{+i\mathbf{q.x}}~. 
        \label{eqn:anni_operator}
\end{eqnarray}
Now, performing Fourier transform of the quark fields and substituting them back in Eq. 
\ref{eqn:anni_operator}, we get the time-evolved annihilation operator $b_{\lambda,s}(t,\mathbf{q})$ entirely in 
terms of time-dependent fermionic wavefunctions $\phi^{u,v}_{\lambda,s}(t^{'},\xs)$. The momentum mode 
occupancy distribution of fermions in a particular momentum state labelled by $\mathbf q$ is defined 
as the expectation value of the number density operator at that particular value of the momentum,

\begin{eqnarray}
    \frac{dN}{d^3 \mathbf{q}}= \frac{1}{2N_c} \sum_{\lambda^\prime,s^\prime} \langle b_{\lambda^\prime,s^\prime}^\dagger(t^\prime,\qs) b_{\lambda^\prime,s^\prime}(t^\prime,\qs) \rangle~.
    \label{eqn:}
\end{eqnarray}
Now, instead of calculating this observable in terms of 
$b_{\lambda^\prime,s^\prime}(t,\mathbf{q}) \text{~and~} b^{\dagger}_{\lambda^\prime,s^\prime}(t,\mathbf{q})$
we can use Eq.~\ref{eqn:anni_operator}  and substitute $b$'s in terms of the 
Fourier modes $u_{\lambda^\prime,s^\prime}(\qs)$ and $\Tilde{\phi}^{u,v}_{\lambda,s}(t^{\prime},\mathbf{q})$, 
the later quantity representing the fermion wavefunctions in the Fourier space,
given as
\begin{equation}
\Tilde{\phi}^{u,v}_{\lambda,s}(t^{\prime},\mathbf{q})=\sum_{\xs} \phi^{u,v}_{\lambda,s}(t^{\prime},\xs)
~\rm{e}^{i \mathbf{q}.\xs}~.
\end{equation}

Thus the momentum distribution can now be re-written as,
\begin{eqnarray}
    \frac{dN}{d^3 \mathbf{q}}= \frac{1}{2N_c}\sum_{\lambda,\lambda^\prime,s,s^\prime} |u^\dagger_{\lambda^\prime,s^\prime}(\mathbf{q}) \Tilde{\phi}^u_{\lambda,s}(t^{\prime},\mathbf{q})|^2~, \nonumber
\end{eqnarray}
where we have used the initial conditions given in Eq. \ref{eqn:initial_cond} to sum over various spin polarizations  
and color quantum numbers.

\textbf{Appendix B: Calculating the second moment of momentum distribution function through fermionic wavefunctions 
and electric field correlators} 

Once we obtain the momentum occupancy distribution at time $t$ from fermionic wavefunctions following the procedure 
described in the earlier subsection, for the whole set of lattice momenta (with $n_x, n_y, n_z$ lattice indices), 
we calculate the second moment (as a function of $\Delta t$) of the distribution and define it as a measure of 
momentum broadening, i.e., 
\begin{eqnarray}
\label{eq:Whatever}
    \frac{\langle q_i^2 \rangle (\Delta t)}{Q^2} 
    &=& \frac{\langle {\hat{q}_i}^2 \rangle (\Delta t)}{(Qa)^2} 
    \;,
\end{eqnarray} where, $\hat{q}_i$ is the lattice momenta defined as,
\begin{eqnarray}
    \hat{q}_i= - \sum_{k} ~ C_k ~\sin \Bigg(k \frac{2\pi n_i}{N}\Bigg)
\end{eqnarray}
where $n_i=0,...,N-1$ is the lattice momentum mode index and $0,1,..,C_{k-1}$ are coefficients used 
for $\mathcal {O}(a^{k})$ improvement of the Wilson-Dirac operator. Specifically, for the NLO i.e. $\mathcal{O}(a^2)$ 
improvement, the coefficients are $C_1=4/3$ and $C_2=-1/6$. Now, from the definition of the second 
moment of the momentum distribution, 
\begin{eqnarray}
    \langle {\hat{q}_i}^2 \rangle (\Delta t) ~ = \sum_{\{n_j\} }{\hat{q}_i}^2 \times ~ \frac{dN}{d^3 \mathbf{q}} ~ \Bigg( \prod_j \Delta \hat{q}_j  \Bigg)
\end{eqnarray}
we can extract $\langle q_i^2 \rangle /Q^2$ as defined in Eq.~\ref{eq:Whatever}. 

Now, to compare the results to those obtained in the \textit{infinite-mass} limit, we need to calculate $\langle q_i^2 \rangle /Q^2$ using color-electric field correlators~\cite{Boguslavski:2020tqz} which is given as,
\begin{eqnarray}
    \frac{\langle q_i^2 \rangle (\Delta t)}{Q^2} = \frac{g^2}{2N_c Q^2} \int_t^{t+\Delta t} dt^\prime \int_t^{t+\Delta t} dt^{\prime \prime} ~\langle E_i^a(t^\prime)E_i^a(t^{\prime \prime)} \rangle \nonumber~,
\end{eqnarray}
where the repeated index $a$ implies a sum over number of color generators $a=1,...,N_c^2-1$. On 
the lattice, the second moment of the momentum distribution can be defined as,
\begin{eqnarray}
    \frac{\langle q_i^2 \rangle (\Delta t)}{Q^2} &=& \frac{1}{2N_c Q^2} ~ \Big{\langle} \Bigg(\sum_{\{t_n \in [t,t+\Delta t]\}} \frac{\hat{E}_i^a (t_n)}{a^2} \times a_t \Bigg)^2 \Big{\rangle}\;, \nonumber \\
\end{eqnarray}
where, $\hat{E}_i^a (t_n)=ga_s^2 E_{a}^{a}(t)$ denote the components of lattice electric field defined 
at time $t_n$ and $a_t$ is the lattice spacing in the temporal direction.

In the diffusive regime, when the time variation of $\frac{\langle q_i^2 \rangle (\Delta t)}{Q^2}$ is linear, 
we can calculate momentum diffusion coefficient $\kappa$ from the second moment of momentum distribution from the following relation,

\begin{eqnarray}
    \frac{\kappa}{Q^3} = \frac{1}{3} \frac{d}{d(Q\Delta t)} \left[\sum_i \frac{\langle q_i^2 \rangle (\Delta t)}{Q^2}\right]~.
\end{eqnarray}
The factor of $1/3$ in the expression of $\kappa/Q^3$ comes from the fact that we have isotropy in the space 
directions and the momentum diffusion coefficient measures the amount of broadening along any of the spatial 
dimensions. We have calculated the slope using $Q\Delta t= 350$ which is long enough for the 
diffusive and hence a linear dependence in time to set in. Values of $\kappa/Q^3$ thus obtained for various 
quark masses and from the color-electric field correlator i.e. in the infinite massive limit are shown in 
table~\ref{tab:kappavsmass}.

\begin{table}[h]
     \begin{tabular}{|c|c|}
         \hline
     Quark mass, $m/Q$ & $\kappa/Q^3$ ($\times 10^{-5}$) \\ [1.5ex]
     \hline
     \hline
     1.2    &  37.13 (5) \\
     \hline
     2.1    &  16.54 (4) \\
     \hline
     4.2    &  8.64 (2)  \\
     \hline
     6.0    &  7.24 (1)  \\
     \hline
     12.0    &  6.20 (2)  \\
     \hline
     \hline
     $\infty$ (EE-correlator) & 5.227 (3) \\
     \hline
    \end{tabular}
    \caption{The momentum diffusion coefficient as a function of quark mass.}
      \label{tab:kappavsmass}
\end{table}

\textbf{Appendix C: Fit to the spectral functions from hard thermal loop perturbation theory} \\

\begin{figure*}[]
    \centering
    \includegraphics[width=\textwidth, height=12cm]{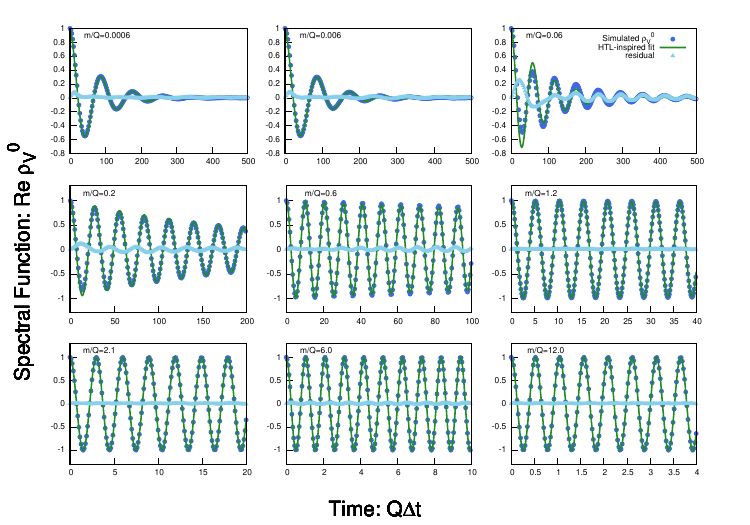}
    \caption{HTL-inspired fit function (green lines) to the $\text{Re }\rho^0_V$ component of the 
    spectral function, shown along with the lattice data (blue points) for a wide range of bare quark mass. In each 
    plot the light-blue points represent the residual i.e. the difference between the fit function and the data.}
    \label{fig:HTLinspiredfit}
\end{figure*}

To extract the effective masses  $m_{\text{eff}}$ 
and damping factor $\gamma$ of quarks we perform a HTL-inspired fit to the 
component of vector spectral function $\text{Re }\rho^0_V$ with the ansatz
$\text{Re }\rho^0_V(Q\Delta t)=~\text{exp}(-\gamma \cdot \Delta t)
~\cos(m_{\text{eff}}\cdot\Delta t)$. In Fig.~\ref{fig:HTLinspiredfit} 
we plot the spectral function measured in our simulations, the fit obtained 
using our ansatz and the residual difference that is obtained subtracting 
the fit function from the data. It is evident from the figure 
that our ansatz for the vector spectral function corresponding 
to light quark masses $m_{\text{bare}}/Q<0.01$ as well as 
heavy quark masses $m_{\text{bare}}/Q \geq 0.6$ provides an 
excellent fit to the data, yielding a quite small residual.
However, for intermediate quark masses for which $m_{\text{bare}} \sim (m_{\text{eff}}-m_{\text{bare}})$ 
the residual, i.e. the difference between the fit-function and data, for the spectral function 
is noticeably larger and we conclude that in this mass range the spectral function is not well described by the aforementioned ansatz.

\bibliography{references.bib}

\end{document}